\documentclass[aps,prd,groupedaddress,notitlepage,nofootinbib,11pt]{revtex4-2}
\usepackage{amsmath,amssymb}
\usepackage{caption}
\usepackage{subcaption}
\usepackage{float}
\usepackage{multirow}
\usepackage{comment}
\usepackage{color}
\usepackage{graphicx}
\usepackage{hyperref}
\makeatletter
\gdef\@fpheader{}
\makeatother

\newcommand\be{\begin{equation}}
\newcommand\ee{\end{equation}}
\newcommand\bea{\begin{eqnarray}}
\newcommand\eea{\end{eqnarray}}

\renewcommand\comment[1]{}

\newcommand\capt[1]{\caption{\textsf{#1}}}

\renewcommand\hat{\widehat}

\begin{document}
\title{Learning holographic horizons}

\author{Vishnu Jejjala}
\email{v.jejjala@wits.ac.za}
\affiliation{Mandelstam Institute for Theoretical Physics, School of Physics, NITheCS, and CoE-MaSS, University of the Witwatersrand, Johannesburg, WITS 2050, South Africa}

\author{Sukrut Mondkar}
\email{sukrutmondkar@hri.res.in}
\affiliation{Harish-Chandra Research Institute, A CI of Homi Bhabha National Institute, Chhatnag Road, Jhunsi, Prayagraj (Allahabad) 211019, India.}
\affiliation{Homi Bhabha National Institute, Training School Complex, Anushakti Nagar, Mumbai 400094,
India}

\author{Ayan Mukhopadhyay}
\email{ayan.mukhopadhyay@pucv.cl}
\affiliation{Instituto de F\'{\i}sica,
Pontificia Universidad Cat\'{o}lica de Valpara\'{\i}so,
Avenida Universidad 330, Valpara\'{\i}so, Chile.}

\author{Rishi Raj}
\email{rishi.raj@lpthe.jussieu.fr}
\affiliation{Laboratoire de Physique Théorique et Hautes Energies -- LPTHE Sorbonne Université and CNRS, 4 Place Jussieu, 75005 Paris, France}

\begin{abstract}
{
We apply machine learning to understand fundamental aspects of holographic duality, specifically the entropies obtained from the apparent and event horizon areas. We show that simple features of only the time series of the pressure anisotropy, namely the values and half-widths of the maxima and minima, the times these are attained, and the times of the first zeroes can predict the areas of the apparent and event horizons in the dual bulk geometry at all times with a fixed maximum length ($10$) of the input vector. We also argue that the entropy functions are the measures of information that need to be extracted from simple one-point functions to reconstruct specific aspects of correlation functions of the dual state with the best possible approximations.  
}
\end{abstract}
%\begin{document}
\maketitle

\section{Introduction}
\label{sec:intro}
The central idea behind modern approaches to Big Data is to devise a predictive analytics by isolating and focusing on salient features within a large dataset.
Often, we do not know what the important features are \emph{a priori}.
Machine learning can be successful at identifying subtle correlations within datasets and extracting relationships in order to make testable predictions.
Unboxing an effective machine learning architecture may in fact teach us about the structure of data.

Spacetime is Big Data.
Moreover, in certain settings, we have prior access to the most efficient data compression protocol available for encoding a dataset.
This is holography~\cite{tHooft:1993dmi, Susskind:1994vu}.
The gauge/gravity correspondence in asymptotically anti-de Sitter (AdS) spaces~\cite{Maldacena:1997re, Gubser:1998bc, Witten:1998qj} and the Matrix model for M-theory~\cite{Banks:1996vh} supply examples where the degrees of freedom of a gravitational system are explicitly in codimension one. In this respect, the fundamental understanding of the emergence of semiclassical spacetime from a strongly coupled gauge theory living at the boundary is also one of the prominent areas of interdisciplinary research in theoretical physics (see~\cite{Harlow:2018fse,Kibe:2021gtw} for reviews).

To date, there have been few investigations of the holographic correspondence in the context of machine learning~\cite{Hashimoto:2018bnb, Hashimoto:2018ftp, Hashimoto:2019bih, Hu:2019nea, Tan:2019czc, Yan:2020wcd, Akutagawa:2020yeo, Hashimoto:2020jug, Song:2020agw, Hashimoto:2021ihd, Lam:2021ugb, Li:2022zjc, Kumar:2023hlu, Hashimoto:2022eij, Katsube:2022ofz, Park:2023slm}. The subject of this paper is to demonstrate that features of the bulk geometry can be machine learned from information contained purely within the dual quantum field theory. In particular, we address the notion of the entropies associated with the dynamical horizons of the emergent geometry corresponding to a typical non-equilibrium state in the dual field theory. This can be expected to provide new insights into some fundamental aspects of the holographic duality since the horizons are among the important gauge invariant features of a spacetime, and the physical understanding of the non-equilibrium entropies associated with the horizons should have important consequences for our understanding of the emergence of spacetime.

We particularly focus on homogeneous thermalization in this work as in this context the Ward identities ensure that the energy density is constant in the absence of any drive. Since the energy density sets the time scale of thermalization, we can understand how the pressure isotropization is encoded at the apparent and event horizons, particularly via the entropy functions by comparing different evolutions with the same constant energy density. Our studies are based on the Chesler--Yaffe method of construction of the dual holographic geometries via numerical relativity~\cite{Chesler:2008hg}. Instead of driving the system via a time-dependent anisotropic boundary metric as in~\cite{Chesler:2008hg}, we set non-trivial initial conditions (which are implicitly generated by homogeneous quenches/driving) corresponding to a fixed energy density in the dual states.

In the language of numerical general relativity, the problem is thus the following. We set initial conditions for a four-dimensional asymptotically anti-de Sitter (aAdS$_4$) geometry which satisfies the vacuum Einstein equation with a negative cosmological constant. The resulting solution of the gravitational equations is dual to a non-equilibrium state which eventually thermalizes, mirroring the ubiquitous feature of gravity that any spacetime eventually settles down to a static AdS$_4$--Schwarzschild black hole. The irreversibility is captured by the growth of the areas of the apparent and event horizons that ultimately coincide, and these give two non-equilibrium entropy functions. From the asymptotic behaviour of the aAdS$_4$ metric, one can read off the (time dependent) energy-momentum tensor in the dual theory, particularly the pressure anisotropy~\cite{Balasubramanian:1999re,deHaro:2000vlm}. 

Here, we show that we can use machine learning to reconstruct the apparent and event horizon areas from simple features of the time series of the pressure anisotropy \textit{without} any explicit knowledge of the initial conditions. Indeed, because there is no independent analytic understanding of the growth of horizon entropies, machine learning is a novel tool for determining how boundary and bulk physics are correlated from a minimal set of inputs.

This problem is also of interest because it has been shown earlier that just with the data of the time dependent event and apparent horizon areas one can explicitly construct simple spacetime metrics of AdS--Vaidya type~\cite{Bhattacharyya:2009uu,Joshi:2017ump} which do not satisfy the Einstein equations but give excellent approximations to many other observables in the dual field theory such as correlation functions.\footnote{An AdS--Vadiya metric has a single non-trivial function, namely, a mass, which is a function of the boundary (field theory) coordinates. An analogue construction is a thermal density matrix with a spacetime dependent temperature. However, the latter is not equivalent to an AdS--Vadiya metric, which is more natural in holography from the point of view of the emergent bulk geometry.} Therefore, deducing the growth of the entropies from simple features of the time series of pressure anisotropy can lead us to predict other observables such as correlation functions of the dual state in holographic theories, while simplifying numerical holography, and opening the door to new insights not only into the dynamics of strongly coupled quantum many-body systems but also into the fundamentals of the holographic duality as discussed in the concluding section.

The plan of the paper is as follows. In Sec. \ref{sec:holo-setup}, we discuss the setup which we use the generate the data for the holographic bulk horizons. In Sec. \ref{sec:ml}, we discuss the training and performance of the neural networks for predictions of the time-dependent areas of the bulk horizons. Finally, in Sec \ref{sec:disc}, we conclude with discussions on the implications of our results.

\section{Holographic setup}
\label{sec:holo-setup}
In the limit of strong coupling and large rank of the gauge group, holography maps the non-equilibrium dynamics of quantum field theories to classical gravitational dynamics in asymptotically AdS spacetimes in one higher spatial dimension~\cite{Maldacena:1997re, Gubser:1998bc, Witten:1998qj}.
In this section, we describe the setup of the dual gravitational problem, which will subsequently be used to generate data for our machine learning task.
To do this, we solve the Einstein equations starting from a non-equilibrium asymptotically AdS metric, which represents the out-of-equilibrium initial state of the dual field theory.
(See~\cite{Chesler:2013lia, vanderSchee:2014qwa} for some excellent reviews on this topic.)
By holographic renormalization~\cite{Balasubramanian:1999re, deHaro:2000vlm, Skenderis:2002wp}, we obtain the expression for the stress tensor of the dual quantum theory.

The bulk metric settles down eventually to a planar AdS--Schwarzschild black hole which is dual to the finite temperature state in the gauge theory. We obtain the entropies by locating the event and apparent horizons in the numerically obtained bulk metrics and computing their time dependent areas.

A typical homogeneous metric of an asymptotically AdS$_4$ black brane in Eddington--Finkelstein coordinates (with flat Minkowski metric on the boundary) is
\begin{equation}\label{general-ads4-metric}
    ds^2 = 2\, dr\, dt - A(r,t)\, dt^2 + S(r,t)^2 \left( e^{-B(r,t)}\, dx^2  + e^{B(r,t)}\, dy^2 \right) ~,
\end{equation}
where $r$ is the bulk radial coordinate (the boundary of spacetime is at $r \rightarrow \infty$ ), $t$ is time coordinate on the boundary, which is a null coordinate in the bulk, and $x$ and $y$ are boundary spatial coordinates.
In what follows, we will denote the bulk coordinates by capital Latin indices and boundary coordinates by Greek indices.
The metric is written in terms of $A$, $B$, and $S$, which are functions of the radial coordinate and time.
We focus on AdS$_4$ as this supplies the simplest examples; it is straightforward to generalize to higher dimensions.

A special case of~\eqref{general-ads4-metric} is the Schwarzschild--AdS$_4$ black brane metric for which
\begin{equation}
A(r,t) = \frac{r^2}{L^2} - \frac{M L^2}{r} ~, \qquad B(r,t) = 0 ~, \qquad S(r,t) = \frac{r}{L} ~,
\end{equation}
where $M$ is the ADM mass~\cite{Arnowitt:1959ah}, which is related to Hawking temperature $T$ by the relation
\begin{equation}
M = \frac{64}{27}L^3 \pi^3 T^3 ~,
\end{equation}
with $L$ being the radius of AdS$_4$.
The event and apparent horizons coincide at the position $r_\text{H} = L^{4/3} M^{1/3}$. We will use units in which $L=1$ since only dimensionless combinations with $L$ relate to physical constants and variables in the dual three dimensional conformal field theory.

The boundary metric is identified with the physical background metric on which the dual field theory lives.  Requiring that it is a codimension one Minkowski space, implies that the functions $A$, $B$, and $S$ should behave as follows\footnote{Here we have fixed a residual gauge freedom by setting the coefficient of $r$ in the expansion of $A(r,t)$ to zero.}
\begin{equation}\label{Eq:Metric-AS}
A(r,t) = r^2 + \frac{a_\text{n}(t)}{r} + \cdots ~, \qquad
B(r,t) = \frac{b_\text{n}(t)}{r^3} + \cdots ~, \qquad
S(r,t) = r - \frac{1}{8 \, r}  + \cdots  ~,
\end{equation}
at large $r$. The normalizable modes $a_\text{n}(t)$ and $b_\text{n}(t)$ give the energy density $\varepsilon$ and the pressure anisotropy ($P_x - P_y$) in the dual state, respectively~\cite{Balasubramanian:1999re, deHaro:2000vlm}. Explicitly, the expectation value of the dual energy-momentum tensor $T^{\mu\nu}$ obtained from holographic renormalization has the following non-vanishing components
\begin{equation}\label{Eq:Dual-EMT}
    \hat{T}^{tt} = \hat\varepsilon = - 2 a_\text{n}(t) ~, \,
    \hat{T}^{xx} = \hat{P}_x= -a_\text{n}(t) - 3 b_\text{n}(t) ~, \,
    \hat{T}^{yy} =  \hat{P}_y =-a_\text{n}(t) + 3 b_\text{n}(t) ~,
\end{equation}
 {where $\hat{T}^{\mu\nu}$ is defined via $T^{\mu\nu} = \kappa^{-1}\hat{T}^{\mu\nu}$ with $\kappa^{-1} = (8\pi G_N)^{-1}$ (in units $L=1$ and $G_N$ being Newton's gravitational constant) identified with $N^2$ up to a numerical constant.} As expected in a conformal field theory, the energy-momentum tensor is traceless. The functions $a_\text{n}(t)$ and $b_\text{n}(t)$ are determined by the initial conditions for the bulk metric as discussed below.

The bulk Einstein equations in the characteristic form are~\cite{Chesler:2013lia}:
\begin{eqnarray}\label{Eq:Char-Eqs}
& \begin{array}{ccc}
 S''= -\frac{1}{4} S \left(B'\right)^2 ~ ~, & \qquad &
 \dot{S}'= \frac{3 S}{2}-\frac{\dot{S} S'}{S} ~ ~,
 \end{array} \label{eq:char} \\
& \begin{array}{ccccc}
 \dot{B}'= -\frac{\dot{S} B'}{S}-\frac{\dot{B} S'}{S} & \qquad &
 A''= \frac{4 \dot{S} S'}{S^2}-\dot{B} B', & \qquad &
 \ddot{S}=\frac{\dot{S} A'}{2}-\frac{\dot{B}^2 S}{4} ~.
 \end{array}
 \nonumber
\end{eqnarray}
Above, the prime denotes the radial derivative, and the dot denotes the derivative along the outgoing radial null geodesic, \textit{e.g.},
\begin{equation}
     \dot{f}(r,t) := \partial_t f + \frac{A(r,t)}{2} f'(r,t) ~.
\end{equation}

The first four expressions in~\eqref{eq:char} are \textit{radial} evolution equations that are solved for the functions $S$, $\dot{S}$, $\dot{B}$, and $A$ in this order at every time step. The last equation for $\ddot{S}$ is a constraint equation implying the conservation of the dual energy-momentum tensor~\eqref{Eq:Dual-EMT} in background Minkowski metric. The latter simply states that $a_\text{n}(t)$ is a constant. 

It should be obvious from the nested characteristic form of the gravitational equations that given the radial profile of $B$ (\textit{i.e.}, $B(r, t=0)$) at an initial time and the initial value of $a_\text{n}$ (which turns out to be a constant), we obtain unique solutions for $S(r,t)$, $B(r,t)$ and $A(r,t)$ satisfying the asymptotic behavior~\eqref{Eq:Metric-AS}, and thus a unique bulk metric. The constraint equation evaluated at several radial locations can be used to monitor the accuracy of numerics.\footnote{For the purpose of numerics, it is useful to further redefine variables so that they take finite values at the boundary.} 

For numerical simulations, it is convenient to use the radial coordinate $z = L^2/r$. We set the range of $z$ between $0$ and $L$ which includes the apparent horizon (and thus the event horizon as well). We also set $L=1$ and $a_\text{n} = -1$ (\textit{i.e.}, the energy density $\hat\varepsilon$ to $2$ at all times).

We use the following family of profiles of $B(z,t=0)$ for specifying the initial conditions:
\begin{align}\label{app-eq:ini-B-prof}
    B(z,t=0)  & = z^3 \left( \mathcal{R}_1 e^{-200 (z - 0.2)^2} + \mathcal{R}_2 e^{-200 (z - 0.5)^2} +  \mathcal{R}_3 e^{-100 (z - 0.8)^2}  + N(z)  \right) \nonumber \\
    N(z)  & = \sum_{i=1}^{50} \mathcal{N}_{1 i} \; \sin \left( \mathcal{N}_{2 i} \; z  +  \mathcal{N}_{3 i} \right)
\end{align}
where $\mathcal{R}_{1,2,3}$ are (pseudo)random numbers between $0$ and $1$ {and $z = L^2/r$}. $\mathcal{N}_{1 i}$ are drawn from a Gaussian distribution with mean zero and standard deviation of $0.01$. $\mathcal{N}_{2 i}$ and $\mathcal{N}_{3 i}$ are (pseudo)random numbers between $0$ and $2 \pi$.
These profiles are superpositions of three Gaussians centered at $0.2$, $0.5$, and $0.8$, respectively, with added Gaussian noise $ N(z)$. By varying the parameters $\mathcal{R}_{1,2,3}$, $\mathcal{N}_{1 i}$, $\mathcal{N}_{2 i}$ and $\mathcal{N}_{3 i}$ with $i = 1, \cdots, 50$, we generate various initial conditions. Thus the total number of these parameters specifying the initial conditions is 153.

We solve the set of nested ordinary differential equations \eqref{Eq:Char-Eqs} in the radial coordinate using the pseudospectral method~\cite{Chesler:2013lia,Boyd}. In the pseudospectral method, the radial coordinate is discretized using a Chebyshev grid with a certain choice of the number of grid points, $n$. In our simulations, we have chosen $n=60$. The time updates are performed using the 4th order Adams--Bashforth method. This method requires knowing the values of the functions to be time evolved at previous four time steps. Therefore, for the first nine time steps, we use smaller time steps and perform numerical integration of the interpolated functions. From the 10th time step onwards, the 4th order Adams--Bashforth method is used for time stepping with the value of time step $\delta t = 0.001$. Finally, the time-dependent pressure anisotropy ($b_\text{n}(t)$ ) in the corresponding state of the dual theory can be readily extracted from the numerical solution.

\subsection{Event and apparent horizons}
In thermal equilibrium, the entropy of the dual field theory is calculated from the area of the event horizon via the Bekenstein--Hawking formula: $S = \frac{A_\text{EH}}{4G}$.
However, in out-of-equilibrium settings, the entropy in the dual field theory can be given by the area of the apparent horizon, which is the outermost trapped surface~\cite{Chesler:2008hg,Booth:2011qy,Engelhardt:2017aux}. The area of the event horizon, which satisfies the second law strictly~\cite{Hawking:1971tu,Wald:1999vt}, enumerates the spacetime degrees of freedom.\footnote{A generalized second law can be argued to hold for general null congruences~\cite{Bousso:1999xy,Bousso:2015mna}.}
The apparent horizon is always interior to (or coincident with) the event horizon.
As well, the apparent horizon is defined locally on a given spacelike slice, whereas the event horizon is a global property of the spacetime, and therefore harder to compute.

The position of the event horizon ($r_\text{EH}$) is obtained from the solution to the equation~\cite{vanderSchee:2014qwa}
\begin{equation}\label{event-horizon-position}
\frac{d }{dt}  r_\text{EH}(t)- \frac{A(r_\text{EH}(t),t)}{2} = 0 ~,
\end{equation}
subject to the boundary condition $r_\text{EH} (t \rightarrow \infty) =  M_f^{1/3}$, where $M_f$ is the ADM mass of the \emph{final} equilibrated black brane.
The position of the apparent horizon ($r_\text{AH}$) is determined by solving the algebraic equation~\cite{vanderSchee:2014qwa}
\begin{equation}\label{apparent-horizon-position}
\dot{S}(r_\text{AH},t) = 0 ~.
\end{equation}
The entropy density of the dual field theory system, which is proportional to the area of either the event or the apparent horizons as appropriate, is given by
\begin{equation}\label{entropy-density}
    s_*(t) = \frac{1}{2 \pi} S(r_*(t),t)^2 ~.
\end{equation}
In the numerical code, we use the variable $z = \frac{1}{r}$ instead of $r$ for the radial bulk coordinate since $z$ has finite range unlike $r$.

{We can use various initial conditions given by $B(z,t=0)$ and $a_\text{n}(t=0)$ and solve the Einstein equations. (Recall that $a_\text{n}(t)$ turns out to be a constant.) Thus we obtain the boundary anisotropy $b_\text{n}(t)$ time series data. Since the dual conformal field theory has no intrinsic scale, we can set $a_\text{n} = -1$ so that the dual energy density is given by $\hat\varepsilon =2$ without any loss of generality. We design the neural network to predict the areas of the event and apparent horizons (at all times) from this boundary anisotropy time series data. }

\section{Machine learning entropy}
\label{sec:ml}

To predict the areas of event and apparent horizons, we employ \emph{deep neural networks}.
(We briefly summarize neural networks in Appendix~\ref{sec:nn}.
See~\cite{Mehta_2019, Ruehle:2020jrk} for excellent contemporary reviews.)
The training data for these predictions are the time series for the boundary pressure anisotropy, $b_\text{n}(t)$.

As mentioned in the previous section, we have used various initial conditions for $B(z,t=0)$ with 153 parameters while holding the energy density fixed (in the appropriate units) to $2$. The latter ensures that all the initial conditions in the dataset thermalize at approximately the same time $t_\text{th} \approx 2$. Since the energy density remains constant throughout the evolution, it sets the final temperature and therefore the time scale for thermalization. After the thermalization time, the event and apparent horizons undergoes a ringdown before coinciding with the horizon of the final static black hole. This ringdown is governed by the quasi-normal mode of the final black hole which also governs the time-dependent pressure anisotropy. Therefore, the areas of the event and apparent horizons, and the pressure anisotropy are sensitive to the details of the initial conditions only prior to the thermalization time. So, it is of interest to understand how the pressure anisotropy encodes the bulk horizon areas only before the thermalization time. 

Although the apparent horizon can be causally determined unlike the event horizon, we are interested only to use discrete time series of the pressure anisotropy for predicting the area of the bulk horizons. Therefore, it is necessary to use the discrete time series of the pressure anisotropy up to a sufficiently long time for the predictions to be sufficiently accurate. We find that the optimal truncation of the discrete time series is $t= t_0 +3$, with $t_0$ being the initialization time, and this is a bit beyond the time $t_0 + t_\text{th} =t_0 +2$ when the solutions thermalize.

We find that the predictions of the event horizon as well as the apparent horizon areas happen almost perfectly at \textit{all times} starting from the initial time \text{both} with the inputs of the \textit{long} and \textit{short} (sparse) data for the time-series of the pressure anisotropy in the dual boundary theory. More precisely, the long and short input vectors are as defined below.

The long input vector is the discrete time series of the pressure anisotropy $b_\text{n}(t_0 + k \delta t)$, where $t_0 =0$ is the initial time, $\delta t=0.01$ and $$k = (10^{-4}, 10^{-7/2}, 10^{-3}, 10^{-5/2}, 10^{-2}, 10^{-3/2}, 10^{-1}, 10^{-1/2}, 1- \sum_{i=-8}^{-1} 10^{i/2})$$ for first nine time steps respectively and $k=1$ from the 10th time step onward. (Although, we use time step $\delta t = 0.001$ for the simulations as mentioned in the previous section to generate elements of our dataset, for the long input vector, we only retain values of various dynamical quantities with $\delta t = 0.01$ after the ninth time stepping.) The full time series is truncated to $t =3$ which is well past the thermalization time scale. 
The \emph{short input vector} contains a handful of only ten features of the pressure anisotropy time series which are
\newpage
\begin{enumerate}
    \item $b_{\text{n}, \text{max}}$: maximum value that $b_\text{n}(t)$ takes;
    \item $t_{\text{max}}$: time $t$ when $b_{\text{n},  \text{max}}$ is attained;
    \item $b_{\text{n},  \text{min}}$: minimum value that $b_\text{n}(t)$ takes;
    \item $t_{\text{min}}$: time $t$ when $b_{\text{n}, \text{min}}$ is attained;
    \item full width at half maximum;
    \item full width at half minimum ;
    \item {times corresponding to first four zeroes of $b_\text{n}(t)$.}
\end{enumerate}
Therefore, the length of the \emph{short input vector} is $10$, which is \textit{significantly} shorter than the number of parameters in the initial condition \eqref{app-eq:ini-B-prof}, which are 153 as discussed in the previous section. Note also that the full time series is truncated to $t =3$.

\begin{figure}[H]
     \centering
     \begin{subfigure}[H]{0.29\textwidth}
         \centering
         \includegraphics[width=\textwidth]{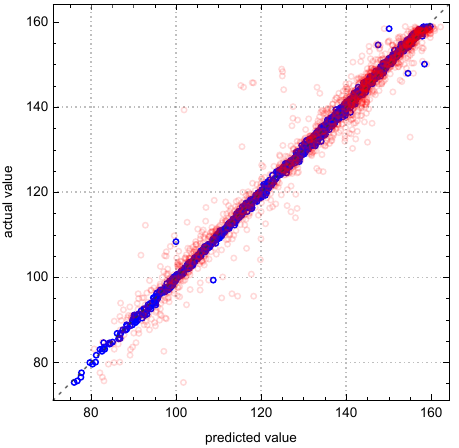}
         \capt{$S_{EH}$ at $t=0$ }
         \label{fig:pred-report-SEH-t=0-short-long}
     \end{subfigure}
     \hfill
     \begin{subfigure}[H]{0.29\textwidth}
         \centering
         \includegraphics[width=\textwidth]{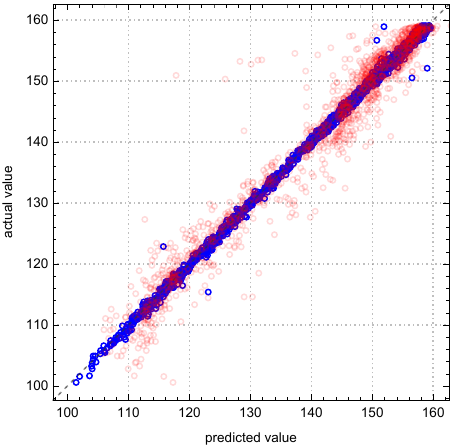}
         \capt{ $S_{EH}$ at $t=0.75$ }
         \label{fig:pred-report-SEH-t=0p75-short-long}
     \end{subfigure}
     \hfill
     \begin{subfigure}[H]{0.29\textwidth}
         \centering
         \includegraphics[width=\textwidth]{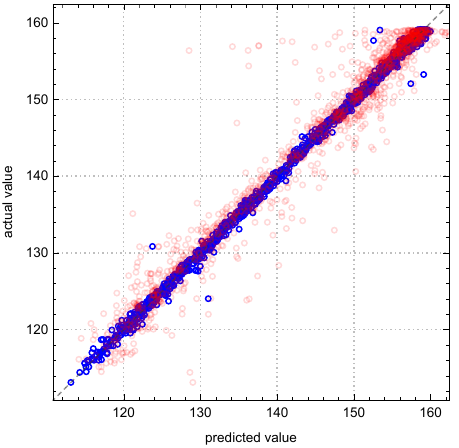}
         \capt{$S_{EH}$ at $t=1$ }
         \label{fig:pred-report-SEH-t=1-short-long}
     \end{subfigure}
     \vfill
     \begin{subfigure}[H]{0.29\textwidth}
         \centering
         \includegraphics[width=\textwidth]{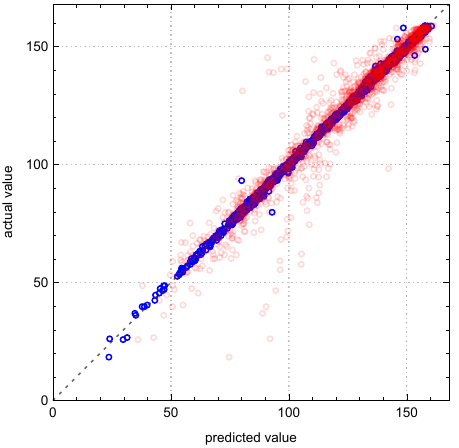}
         \capt{$S_{AH}$ at $t=0$ }
         \label{fig:pred-report-SAH-t=0-short-long}
     \end{subfigure}
     \hfill
      \begin{subfigure}[H]{0.29\textwidth}
         \centering
         \includegraphics[width=\textwidth]{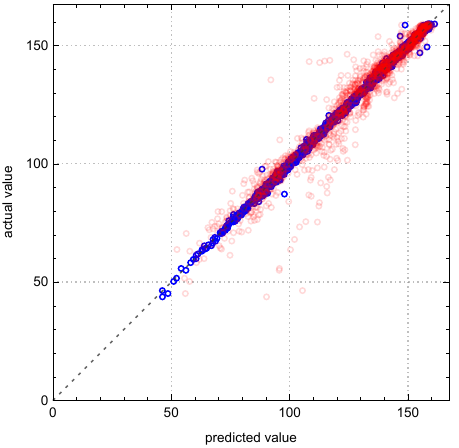}
         \capt{$S_{AH}$ at $t=0.75$ }
         \label{fig:pred-report-SAH-t=0p75-short-long}
     \end{subfigure}
      \hfill
      \begin{subfigure}[H]{0.29\textwidth}
         \centering
         \includegraphics[width=\textwidth]{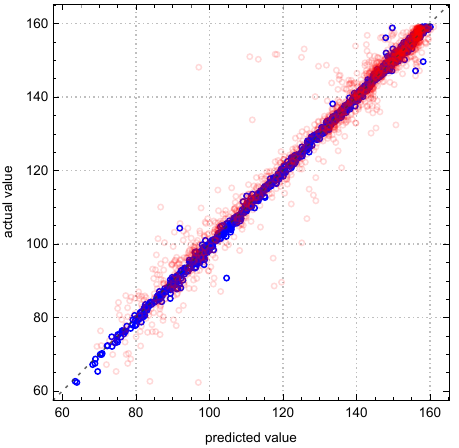}
         \capt{$S_{AH}$ at $t=1$ }
         \label{fig:pred-report-SAH-t=1-short-long}
     \end{subfigure}
        \capt{Testing statistics for $S_\text{EH}$ and $S_\text{AH}$ with both \textbf{long input vector} (blue) as well as \textbf{short input vector} (red) at various times. Note that the scale is set by the (conserved) energy density which is the same for all the elements of the dataset used for machine learning, which makes it meaningful to compare the times across the dataset. 
        }
        \label{fig:testing-SEH-SAH-short-long}
\end{figure}

Our main result is that training the neural network with only the handful of simple features of $b_\text{n}(t)$ time series in the short input vector suffices to generate predictions comparable to the predictions made by the neural network trained on the almost full $b_\text{n}(t)$ time series data, namely the long input vector. 
The neural network predicts both the event and apparent horizon areas (almost) perfectly at all times in both cases. Fig.~\ref{fig:testing-SEH-SAH-short-long} gives representative plots for the predictions of the neural network for the event and apparent horizon areas at initial time, and at times 0.5 and 1 after the initial time, both when it is trained with the long and short input vectors, respectively.

To assess the performance of the neural network, we compute the following quantities over the test set:
\begin{eqnarray}
\text{mean deviation} \;  (\text{MD}): && \text{Mean}[\frac{\text{Abs}[\text{actual} - \text{predicted}]}{\text{actual}}] ~, \label{eq:accuracy} \\
\text{baseline}: && \text{Mean}[\frac{\text{Abs}[\text{actual} - \text{Mean}[\text{actual}]]}{\text{actual}}] ~. \label{eq:baseline}
\end{eqnarray}
The baseline~\eqref{eq:baseline} simply assigns the mean in the dataset as a universal prediction.
The accuracy of the trained neural network is compared to this baseline.
We as well calculate the standard deviation (SD) between the actual value and the predicted value, $R$-squared, and $\chi$-squared in the usual way. These statistics are reported in Tables~\ref{tab:SEH} and~\ref{tab:SAH}.

\comment{
\begin{align*}
    \text{MD} =& \text{Mean}[\frac{\text{Abs}[\text{actual}] - \text{Abs}[\text{predicted}]}{\text{actual}}]; \quad  \text{DFM} = \text{Mean}[\frac{\text{Abs}[\text{actual} - \text{Mean}[\text{actual}]]}{\text{actual}}]; \\ \text{SD} =& \text{Standard Deviation}; \quad \chi\text{-squared} = \sum \frac{(\text{actual} - \text{predicted})^2}{\text{actual}}
\end{align*}
}

\begin{table}[H]
    \centering
\begin{tabular}{ |c|c|c|c|c|c|c| } 
\hline
time & input vector & MD & baseline & SD & $R$-squared & $\chi$-Squared \\
\hline
\multirow{2}{4em}{$t=0$} & long & 0.00422111 & 0.14546 & 0.818 &  0.998 &  7.77747 \\ 

& short & 0.018707 & 0.14546 & 3.91  & 0.964 & 187.567 \\ 
\hline
\multirow{2}{4em}{$t=0.75$} & long & 0.00323828 & 0.101198 & 0.686  & 0.998 &  5.1811 \\ 

& short & 0.0159789 & 0.101198 & 3.41  & 0.952 & 128.494 \\ 
\hline
\multirow{2}{4em}{$t=1$} & long & 0.00276327 & 0.0821333 & 0.617 & 0.998 &  4.063 \\ 

& short & 0.0119103 & 0.0821333 & 2.85 & 0.952 & 85.4856 \\
\hline
\end{tabular}
\capt{Various statistical measures assessing performance of neural network for $S_\text{EH}$ prediction.}
    \label{tab:SEH}
\end{table}

\begin{table}[H]
    \centering
\begin{tabular}{ |c|c|c|c|c|c|c| } 
\hline
time & input vector & MD & baseline & SD & $R$-squared & $\chi$-Squared \\
\hline
\multirow{2}{4em}{$t=0$} & long & 0.00681106  & 0.246200 & 1.10 & 0.998 & 19.3582 \\ 

& short & 0.0452788 & 0.246200 & 7.54 & 0.930 & 1262.02 \\ 
\hline
\multirow{2}{4em}{$t=0.75$} & long & 0.00582296 & 0.194025 & 1.02 & 0.998 & 13.5395  \\ 

& short & 0.0326038  & 0.194025 & 6.02 & 0.942 & 627.272 \\ 
\hline
\multirow{2}{4em}{$t=1$} & long & 0.0055171 & 0.180533 & 1.06  & 0.998 & 15.0186 \\ 

& short & 0.0227724  & 0.180533  & 4.71 & 0.962 & 289.725 \\ 
\hline
\end{tabular}
\capt{Various statistical measures assessing performance of neural network for $S_\text{AH}$ prediction.}
    \label{tab:SAH}
\end{table}

We have performed a more robust test of the performance of the neural networks with respect to the predictions when trained with short or long input vectors via \emph{$k$-fold cross validation}. This has been reported in Appendix \ref{sec:k-fold}.

\section{Discussion and outlook}
\label{sec:disc}
In this work, we have demonstrated that neural networks can learn some aspects of bulk geometry given simply the time series of a one-point function (expectation value of an operator) as input. 
In particular, we have established that the areas of the event and apparent horizons can be learned from the time series data of the pressure anisotropy in the boundary field theory. Furthermore, neural networks also predict the black hole entropies at all times from only a small number of features of the time series data, namely the values and half-widths of the maxima and minima, the times the latter are attained, and the times of the first zeroes with the maximum length of the input vector set to $10$ and with the time series truncated to a maximum value of time commensurate with the thermalization time. 

Generally, in order to construct the bulk spacetime metric, one would need the boundary data and also initial conditions. In our work, we have considered a wide class of initial conditions, and we have demonstrated that machine learning of the event and apparent horizon entropies is possible with inputs of \textit{only} compressed boundary data.  These results are surprising and point to universality classes of bulk metrics and their dual states which can be characterized only by simple features of the time series of simple observables, namely one-point functions. Our work therefore indicates that neural networks are not only promising tools for addressing bulk reconstruction in AdS/CFT but also for obtaining new insights into holographic dynamics.

We argue that our approach can yield the holographic interpretation of the entropy functions obtained from the evolving event and apparent horizons in dynamical black hole geometries dual to non-equilibrium states. It is natural to think of entropy functions as determining simplistic statistical ensembles which can be constructed from simple inputs and which provide best approximations to all observables of a given state. In fact, a precise holographic interpretation of semiclassical spacetime itself as a minimum bias statistical ensemble has been advocated in~\cite{Jafferis:2022wez}. The event and apparent horizon entropies can be naturally thought of as further simplifications of the emergent spacetime. 

The first step towards developing such a precise statistical interpretation of non-equilibrium event and apparent horizon entropies would be to test if machine learning of these entropy functions is possible without a detailed knowledge of the two-derivative gravity theory (Einstein's theory coupled to a few bulk fields) governing the bulk dynamics in holographic duality. If the entropies can lead to a statistical ensemble which can approximate observables of the non-equilibrium state dual to the dynamical black hole geometry, the algorithm for obtaining such an ensemble should not depend on the specific details of the theory. Conversely, it should be possible to extract these entropy functions without detailed knowledge of the dual strongly interacting holographic theory.

Indeed the results of~\cite{Joshi:2017ump} strongly suggest that the entropy functions can be learned from simple boundary data without detailed knowledge of the underlying microscopic theory or the dual bulk gravitational theory. It was shown in~\cite{Joshi:2017ump} that a simple AdS-Schwarzschild geometry with a time-dependent mass (\textit{viz.}, an AdS-Vaidya geometry)  gives the \textit{best} approximation to non-equilibrium causal correlation functions provided it reproduces the \textit{exact} time-dependent apparent horizon entropy of the \textit{exact} dual gravitational solution dual to the given state. Similarly, a simple  Ads--Vaidya geometry gives the \textit{best} approximation to coarse-grained features of non-equilibrium causal correlation functions provided it reproduces the \textit{exact} time-dependent event horizon entropy of the \textit{exact} dual gravitational solution dual to the given state\footnote{This work build on earlier suggestions of \cite{Bhattacharyya:2009uu} where AdS-Vaidya geometries were proposed as the starting point of approximating slow quenches. However, instead of giving exact event/apparent horizon entropies, these AdS-Vaidya geometries were supposed to reproduce exact one-point functions like pressure anisotropies. In~\cite{Joshi:2017ump}, it was shown that better approximations are obtained if the AdS-Vaidya geometries are constructed such that exact event/apparent horizon entropies are reproduced. The numerical studies in~\cite{Joshi:2017ump} were based on methods developed in~\cite{Banerjee:2016ray}.}. In both cases, the approximations are typically better than one percent on average. 

Such AdS-Vaidya geometries are natural gravitational analogues of thermal density matrices with just time-dependent temperatures (although not exactly equivalent to the latter) as their construction require only inputs of the entropy functions (and no knowledge of the dual gravitational theory). Our present work together with the results of \cite{Joshi:2017ump} suggest that conversely the entropy functions can possibly be learned from simple one-point functions \textit{without} knowledge of the dual bulk two-derivative gravitational theory, and thus such entropy functions can be interpreted as measures of information which need to be extracted from one-point functions in order to construct quantum statistical ensembles that give best possible approximations to the dual states, especially for reproducing the two-point correlation functions. Our approach can be useful to arrive at such a precise interpretation.

In the future, we would like to unbox the neural network in order to understand how learning takes place, for instance, by employing techniques like layer-wise relevance propagation~\cite{montavon2019layer, Craven:2020bdz} and this should be important for the stated goal of interpreting the entropy functions. Because this work is a proof of principle that we can apply Big Data techniques to holography and machine learn salient features of spacetime dynamics from dual data, we do not have a need to optimize the network hyperparameters to improve performance. It would be good to do this, of course.
\section*{Acknowledgements}
We are grateful to Jessica Craven, Koji Hashimoto, Shivaprasad Hulyal, Dileep Jatkar, Lata Joshi, Arjun Kar, Tanay Kibe, David Mateos, and Harald Skarke for enlightening discussions on this and related work. We acknowledge the High-Performance Scientific Computing facility of the Harish-Chandra Research Institute, where we generated most of the data used for training and testing our neural networks.
We thank the organizers and participants of String Data 2022, where aspects of this work were presented.
VJ is supported by the South African Research Chairs Initiative of the Department of Science and Innovation and the National Research Foundation.
VJ would as well like to thank the Isaac Newton Institute for Mathematical Sciences for support and hospitality during the program ``Black holes: bridges between number theory and holographic quantum information'' during which work on this paper transpired; this work was supported by EPSRC grant number EP/R014604/1.
%SM is supported by .... 
The research of AM was partly supported by the center of excellence grants of the Ministry of Education of India. 
%RR is supported by ....

\appendix

\section{Feedforward neural networks}\label{sec:nn}
Consider the map
\begin{equation}
\begin{array}{cccc}
f_i: & \mathbb{R}^{n_{i-1}} & \quad \to \quad & \mathbb{R}^{n_i} \cr
& \mathbf{v}_{i-1} & \quad \mapsto \quad & \mathbf{v}_i = \sigma(\mathbf{v}_i')
\end{array} ~,
\end{equation}
where $\mathbf{v}_i' = W^{(i)} \cdot \mathbf{v}_{i-1} + \mathbf{b}_i$.
The vector $\mathbf{b}_i \in \mathbb{R}^{n_i}$ is called a \emph{bias vector} and the $n_i\times n_{i-1}$ matrix $W^{(i)}$ is called a \emph{weight matrix}.
The \emph{activation function} $\sigma$ acts non-linearly elementwise on the components of the vector $\mathbf{v}_i'$.
In our case, we choose $\sigma(x)=x\,\Theta(x)$, where $\Theta(x)$ is the Heaviside step function.
(This is the rectified linear unit or ReLU activation.)
Taking the vector $\mathbf{v}_\text{in} := \mathbf{v}_0 \in \mathbb{R}^{n_0}$ as the input vector, this procedure is iterated:
\begin{equation}
\mathbf{v}_\ell = f_\ell(f_{\ell-1}(\cdots(f_1(\mathbf{v}_0)\cdots))) ~, \qquad
a_\text{out} := \sum_{m=1}^{n_\ell} \mathbf{v}_\ell^m ~.
\label{eq:nn}
\end{equation}
In writing $a_\text{out}$, we have summed the components of the vector $\mathbf{v}_\ell$.

Suppose we have a set of input vectors labeled by $j$ that we use for training.
To each of these input vectors $\mathbf{v}_{\text{in},j}$ we associate a target value $\alpha_j \in \mathbb{R}$.
We then consider a mean squared loss function that compares each $\alpha_j$ to its predicted value $a_{\text{out},j}$.
Backpropagation enables us to tune the elements of the weight matrices and bias vectors to minimize this loss.
In this way, we have constructed an $\ell$-hidden layer fully connected feedforward neural network that is trained to approximate $\alpha_j$ from the features encoded in $\mathbf{v}_{\text{in},j}$ via stochastic gradient descent.
The dimension $n_i$ specifies the number of neurons in the $i$-th hidden layer.

In our experiments, the architectures of the neural network are dependent on the vector $\mathbf{v}_\text{in}$.
{With long input vectors in $\mathbb{R}^{299}$, we use three hidden layers with $600$ neurons each.
With short input vectors in $\mathbb{R}^{10}$, we use three hidden layers with $20$ neurons each.}
(The input vectors are described in Sec. \ref{sec:ml}.)
We use $70\%$ of a $10000$ data points for training, $15\%$ for validation, and $15\%$ for testing.
Performance statistics are reported on the test set.
The neural networks are implemented in \texttt{Mathematica}~\cite{Mathematica} and use \textsf{Adam} for optimization.
We make our code available on \textsf{Github}~\cite{GitHub-ml-holo-23}.

\comment{
\begin{table}[]
    \centering
    \begin{tabular}{|c|c|c|c|c|}
    \hline
         Input Vector & Input Vector Length & Neurons & Architecture \\
         \hline 
        Long & 1000 & 2000 & Linear, ReLU, Linear, ReLU, Linear \\
        \hline
        Short & 100 & 300 & Linear, ReLU, Linear, ReLU, Linear \\
        \hline
    \end{tabular}
    \capt{Details of NN}
    \label{tab:nn-arch}
\end{table}
}

%\section{Data preparation and analysis}
%\label{sec:app-1}

\section{$k$-fold cross validation}
\label{sec:k-fold}

As a further test of the robustness of predictive power of our neural networks trained on both long and short input data, we perform \emph{$k$-fold cross validation}. In this method, one randomly divides the entire dataset into $k$ subsets (also called $k$ folds). Then, one keeps one of these $k$ folds as testing data and trains the neural network on the remaining $k-1$ folds. One repeats this $k$ times, each time choosing a different fold as the testing dataset and the corresponding remaining $k-1$ folds as the training dataset. Finally, one takes the average and standard deviation of all $k$ runs' learning statistics. The most common choices of $k$ are $5$ or $10$. We chose $k=5$.

We find that the various statistical measures reported in tables \ref{tab:SEH} and \ref{tab:SAH} don't vary much across $5$-folds. In particular $R$-squared stays $\sim 1$ and others also don't deviate significantly from their mean value across $5$-folds which are close to their corresponding  values reported in tables \ref{tab:SEH} and \ref{tab:SAH}. We report the average values of each of the quantities of  tables \ref{tab:SEH} and \ref{tab:SAH} across $5$-folds and also their standard deviation (SD) in the tables \ref{tab:SEH-$5$-fold} and \ref{tab:SAH-$5$-fold}.

\begin{table}[H]
    \centering
\begin{tabular}{ |c|c|c|c|c|c|c|c| } 
\hline
 \multicolumn{2}{|c|}{} & \multicolumn{2}{|c|}{$t=0$} & \multicolumn{2}{|c|}{$t=0.75$} & \multicolumn{2}{|c|}{$t=1$} \\ 
 \cline{3-8}
\multicolumn{2}{|c|}{} & long & short  & long & short  & long & short \\
\hline
\multirow{5}{3.5 em}{ Average} & MD & 0.0053267 & 0.022483  &  0.0045241	& 0.017409 & 0.0046980 & 0.012438 \\ 
\cline{2-8}

 & baseline & 0.14975 &  0.14975 & 0.10403 & 0.10403 & 0.08427 & 0.08427 \\ 
\cline{2-8}
  & SD & 1.1996 & 4.836 & 1.0022 & 3.91 & 1.04 & 3.074 \\ 
\cline{2-8}
   & $R$-squared & 0.9952 & 0.9432 & 0.9944 & 0.9358 & 0.993 & 0.9458 \\ 
\cline{2-8}
    &  $\chi$-Squared & 33.345  & 431.96 & 20.283 & 241.93 & 18.168 & 136.94 \\ 
\hline
\multirow{5}{3.5 em}{SD } & MD & 0.0040222  & 0.0052088 & 0.0024334 & 0.0035442 & 0.0018669 & 0.0010649 \\ 
\cline{2-8}

 & baseline & 0.0024558 & 0.0024558 & 0.0014224 & 0.0014224 & 0.0011119 & 0.0011119 \\ 
\cline{2-8}
  & SD &  0.88969 & 1.3000 & 0.65153 & 0.92081 & 0.46326 & 0.36657 \\ 
\cline{2-8}
   & $R$-squared & 0.0061400 &  0.031885 & 0.0056833 & 0.030078 & 0.0064420 & 0.013442 \\ 
\cline{2-8}
    &  $\chi$-Squared & 38.258  & 229.85 & 19.683 & 110.71 & 15.443 & 32.451 \\ 
\hline
\end{tabular}
\capt{Average and standard deviation (SD) across $5$-folds (of cross validation) of various statistical measures assessing performance of neural network for $S_\text{EH}$ prediction}
    \label{tab:SEH-$5$-fold}
\end{table}

\begin{table}[H]
    \centering
\begin{tabular}{ |c|c|c|c|c|c|c|c| } 
\hline
 \multicolumn{2}{|c|}{} & \multicolumn{2}{|c|}{$t=0$} & \multicolumn{2}{|c|}{$t=0.75$} & \multicolumn{2}{|c|}{$t=1$} \\ 
 \cline{3-8}
\multicolumn{2}{|c|}{} & long & short  & long & short  & long & short \\
\hline
\multirow{5}{3.5 em}{ Average} & MD &  0.0129314 & 0.040043 & 0.00998527 & 0.032521 & 0.00753043 & 0.029117 \\ 
\cline{2-8}

 & baseline & 0.25751  & 0.25751 & 0.20153 & 0.20153 & 0.18673 & 0.18673 \\ 
\cline{2-8}
  & SD & 1.6724 & 6.702 & 1.3656 & 5.998 & 1.0708 & 5.804 \\ 
\cline{2-8}
   & $R$-squared & 0.9952  & 0.9466 & 0.9954 & 0.9438 & 0.9972 & 0.9434 \\ 
\cline{2-8}
    &  $\chi$-Squared &  71.6847 & 1085.3 & 48.1872 & 754.47  & 28.3094 & 644.71 \\ 
\hline
\multirow{5}{3.5 em}{SD } & MD & 0.01067  & 0.0044043 & 0.00881007 & 0.0036941 & 0.00654729 & 0.0043984 \\ 
\cline{2-8}

 & baseline &  0.010400 & 0.010400  & 0.0050024 & 0.0050024 & 0.0029318 & 0.0029318 \\ 
\cline{2-8}
  & SD &  1.31589 & 0.73291 & 1.16507 & 1.0039 & 0.777636 & 0.98571 \\ 
\cline{2-8}
   & $R$-squared & 0.00535724  & 0.012177 &  0.00650385 & 0.020535 & 0.00363318 & 0.020598 \\ 
\cline{2-8}
    &  $\chi$-Squared & 77.777  & 325.87 & 65.9824 & 256.38 & 39.0392 & 215.72 \\ 
\hline
\end{tabular}
\capt{Average and standard deviation (SD) across $5$-folds (of cross validation) of various statistical measures assessing performance of neural network for $S_\text{AH}$ prediction}
    \label{tab:SAH-$5$-fold}
\end{table}

 %\bibliographystyle{elsarticle-num} 
 \bibliographystyle{utphys}
 \bibliography{cas-refs}





% \end{thebibliography}
\end{document}